\documentclass[prb,aps,twocolumn,showpacs]{revtex4}

\usepackage{graphicx}
\usepackage{dcolumn}
\usepackage{epsfig}
\usepackage{latexsym}
\usepackage{amsfonts}
\usepackage{amsmath, amsthm, amssymb}
\usepackage{multirow}
\usepackage{booktabs}
\usepackage{appendix}
\usepackage{color}

\newcommand {\otoprule }{\midrule [\heavyrulewidth]}

\begin{document}

\title{Exact Ground State Energy of Hubbard Rings in the Atomic Limit}

\author{W. B. Hodge}

\author{N. A. W. Holzwarth}

\author{W. C. Kerr}
\affiliation{Department of Physics, Wake Forest University,
Winston-Salem, North Carolina 27109-7507}

\date{\today}

\begin{abstract}
Using a straightforward extension of the analysis of Lieb and Wu,
we derive a simple analytic form for the ground state energy of a
one-dimensional Hubbard ring in the
atomic limit. This result is valid for an \textit{arbitrary} number of
lattice sites $L$ and electrons $N \leq L$. Furthermore, our analysis, including
an application of the theory of stochastic matrices, provides insight into the
degeneracy and spin properties of the ground states in the atomic limit.
We give numerical results which illustrate how the atomic limit is approached.

\end{abstract}


\maketitle

\section{Introduction \label{intro}}
For nearly fifty years, the Hubbard model \cite{Hubbard:1963}
has been used  to describe many-body effects in
solids,  capturing  the dominant competition between the
delocalizing effects of the kinetic energy (with strength
described by a hopping energy $t$) and the localizing effects of
the electron-electron repulsion (with strength described by
an
on-site Coulomb energy $U$).
Despite its simple form, the Hubbard model
 has provided significant insight into many-body properties of solids
such as
metal-insulator transitions, high-temperature superconductivity,
and magnetic states,\cite{Marder:2001}
largely because of the accessibility of its analytic and numerical solutions.

The analytic understanding of the Hubbard model stems primarily from
the seminal work of Lieb and Wu\cite{Lieb:1968,Lieb:2003} who derived a
``Bethe anzatz" method\cite{Bethe:1931} for
determining  eigenvalues and eigenfunctions of the single band, one-dimensional
Hubbard model with $L$ lattice sites and $N$ electrons.   They obtained
an explicit expression for the ground state energy for a half-filled system
in the thermodynamic limit ($N = L \rightarrow \infty$). In the intervening
years, many authors have extended the Lieb-Wu analysis. For example,
Essler, Korepin, and Schoutens\cite{Essler:1991,Essler:1992} have shown
that for even $L$ and periodic boundary conditions,
it is possible to find operators to extend the
Bethe ansatz solutions and to find a complete set of eigenstates for the
model.  Lieb and Wu\cite{Lieb:2003} and
Goldbaum\cite{Goldbaum:2005} have presented some existence proofs for
Bethe ansatz solutions for ground states. In addition, many authors have applied the
Lieb-Wu equations to particular systems.\cite{Eggarter:1979, WoynarovichIbid:1982, Ogata:1990}

In the present paper, we reanalyze the Lieb-Wu solutions to the single band, one-dimensional
Hubbard model for the case that there are $L$ lattice sites ($L \le \infty$)
with periodic boundary conditions and $N$ electrons ($N \le L$) in the
limit $U/t \equiv u \rightarrow \infty$. This represents the strongly correlated
regime of the system for which we find that the exact ground state energy can be expressed by
the simple analytic form
\begin{equation} \label{ground state energy}
    E^{}_g = -2t \frac{\sin \left( \displaystyle \frac{\pi N}{L}
    \right)}{\sin \left( \displaystyle \frac{\pi}{L} \right)} .
\end{equation}
To the best of our knowledge, this exact result
has not appeared in the literature as the $u \rightarrow \infty$
ground state energy of  the one-dimensional
Hubbard model with periodic boundary conditions
for general values of $L$ and $N \le L$.  On the other hand,
there are some  closely related
results,\cite{Caspers:1989,Kotrla:1990,Schadschneider:1995,Kumar:2009} and it
is consistent with the results derived for the thermodynamic limit.\cite{Shiba:1972,Doucot:1989, Mielke:1991}
The simple analytic form (\ref{ground state energy}) is helpful for analysis
and numerical studies of the one-dimensional Hubbard model and its derivation
provides insight into the nature of the eigenstates of
the one-dimensional Hubbard model.

In Sec. (\ref{deriv}) we present the derivation of the ground state energy
expression (\ref{ground state energy}). The analysis, which requires a simple  extension of the
Lieb-Wu approach,  helps to illuminate properties of
the model. In Sec. (\ref{illus}) we show several
examples of  systems with fixed $L$ and $N$, illustrating
how total energy eigenvalues converge from their
finite to infinite $u$ values. In Sec. (\ref{discuss}) we discuss the consequent properties of the ground
state wavefunctions for this system and summarize our results
 in the context of the literature.

\section{Derivation of ground state energy \label{deriv}}
The Hamiltonian of the Hubbard model is \cite{Hubbard:1963}
\begin{eqnarray}
    \mathcal{H}
    & = &
    \mathcal{H}_{\mathrm{hop}}(t) + \mathcal{H}_{\mathrm{int}}(U) \nonumber \\
    & = &
    -t \sum_{\langle q, r \rangle} c^{\dag}_{q ,
    \sigma} c^{}_{r, \sigma} + U \sum_{q} n^{}_{q, \uparrow}
    n^{}_{q, \downarrow} , \label{Hubbard Hamiltonian}
\end{eqnarray}
where $c^\dag_{q, \sigma}$ $(c^{}_{q, \sigma})$ creates
(annihilates) an electron with spin $\sigma$ in the Wannier state
localized at lattice site $q$, and $\langle q, r \rangle$ denotes
that a sum is over nearest neighbor sites only.  For the case
of a one-dimensional system with periodic boundary conditions,
the site index $q$  (or $r$) takes values $1 \le q \le L$ and
indices $q$ and $q+L$ are equivalent.



Lieb and Wu considered solutions to the one-dimensional Hubbard model for
a specified electron spin distribution with
$N_\uparrow$ and $N_\downarrow$ indicating the total number of up
and down $z$-component spin orientations, respectively, where
$N = N_\uparrow + N_\downarrow$.
Applying the Bethe ansatz, the total energy eigenvalues are written in a
form identical to that for independent electrons \cite{Lieb:1968}
\begin{equation} \label{Bethe ansatz energy}
    E(N_{\downarrow},N_{\uparrow}) = -2t \sum_{j = 1}^{N} \cos k_j .
\end{equation}
The so-called charge momenta $k_j$, however, are \textit{not} the
wavevectors one finds in the absence of electron-electron
interactions. To find the charge momenta requires solving the set
of coupled nonlinear equations (the Lieb-Wu equations)
\begin{equation} \label{Lieb and Wu equation 1}
    L k_j(N_{\downarrow},N_{\uparrow})
 = 2 \pi I_j + 2 \sum_{\beta = 1}^{N_\downarrow}
    \tan^{-1} \left[ \frac{4}{u}
    \left( \sin k_j - \lambda_\beta \right) \right],
\end{equation}
and
\begin{align} \label{Lieb and Wu equation 2}
    2 \sum_{j = 1}^{N} \tan^{-1} & \left[ \frac{4}{u} \left(
    \lambda_{\alpha} - \sin k_j \right) \right] = 2 \pi J_{\alpha}
    \nonumber \\
    & + 2 \sum_{\beta = 1}^{N_\downarrow} \tan^{-1} \left[ \frac{2}{u} \left(
    \lambda_{\alpha} - \lambda_{\beta} \right) \right].
\end{align}
In this formulation,
Lieb and Wu assume  $N \le L$ and $N_\downarrow \le N_\uparrow$.   The parameter
$I_j=I_j(N_\downarrow)$ is an integer (half-odd-integer) if $N_\downarrow$ is
even (odd), where $1 \le j \le N$.   The
parameter  $J_\alpha=J_\alpha(N_\uparrow)$ is an integer (half-odd-integer) if
$N_\uparrow = N - N_\downarrow$ is odd (even), where
$1 \le \alpha \le N_\downarrow$.
 The $\lambda 's$ are a set of ordered, unequal
real numbers $\lambda_1 < \lambda_2 < ... <
\lambda_{N_\downarrow}$.
 Details of
the derivation of these equations are given in Yang's examination
of a one-dimensional system with delta function interaction
\cite{Yang:1967} as well as a more recent review of the Hubbard
model by Lieb and Wu.\cite{Lieb:2003}


In the present paper, we focus on the limit $u \rightarrow \infty$, henceforth referred to
as the atomic limit. This limit, which represents
a highly correlated system, simplifies the mathematical properties of the Hubbard model
considerably.  In particular, since we are considering the case where
 $N \leq L$, it is reasonable to assume that the charge momenta for the ground state
are real so that
terms of the form $(\sin k_j)/u$ in Eqs.\
(\ref{Lieb and Wu equation 1}) and (\ref{Lieb and Wu equation 2})
vanish in the atomic limit. This assumption is not valid if
$N > L$ since the ground state energy is linear in $u$, making the $k_j$'s necessarily complex.\cite{WoynarovichIbid:1982} As a
result, Eq.\ (\ref{Lieb and Wu equation 2}) simplifies to
\begin{equation}
    2 N \tan^{-1} \left( \frac{4 \lambda_\alpha}{u} \right) = 2
    \pi J_{\alpha} + 2 \sum_{\beta = 1}^{N_\downarrow} \tan^{-1} \left[
    \frac{2}{u} \left( \lambda_\alpha - \lambda_\beta \right)
    \right] .
\end{equation}
By substituting this expression into Eq.\ (\ref{Lieb and Wu
equation 1}) we obtain an equation for the charge momenta at $u =
\infty$,
\begin{equation} \label{k_j}
    k_j(N_\downarrow, N_\uparrow) = \frac{2 \pi}{L} \left[ I_j + \frac{1}{N} \sum_{\beta}^{N_\downarrow}
    J_\beta \right] .
\end{equation}

In order to analyze the ground state, one
possibility\cite{Lieb:1968,Lieb:2003} is to choose
 $I_j$ and $J_\alpha$ to be consecutive integers (or half-odd-integers) centered around the
origin.  We will consider other possibilities later.   With this choice,
 if $N$ is even,
\begin{equation}
    \sum_{\beta=1}^{N_\downarrow} J_\beta = 0 ;
\end{equation}
otherwise, if $N$ is odd,
\begin{equation}
    \sum_{\beta=1}^{N_\downarrow} J_\beta = \frac{N_\downarrow}{2}
    .
\end{equation}
It thus follows that the charge momenta have the following relationships
\begin{equation} \label{Bethe k_j}
    k_j(N_\downarrow, N_\uparrow) = \left\{
    \begin{array}{ll}
        \displaystyle \frac{2 \pi I_j}{L} , & \text{$N$ even}
        \\ \\
        \displaystyle \frac{2 \pi}{L} \left( I_j + \frac{N_\downarrow}{2N}
        \right) , & \text{$N$ odd.}
    \end{array}
    \right.
\end{equation}

For a given system of $L$ sites and $N \le L$ electrons,
we now show that the lowest energy obtained by substituting the
charge momenta given in Eq.\ (\ref{Bethe k_j}) into the energy
expression given in Eq.\ (\ref{Bethe ansatz energy}) is not
necessarily the energy of the ground state. To do so, we present some
examples where this is not the case.

First, we consider
a configuration with all electrons having the same spin ($N_{\downarrow},
N_{\uparrow}) = (0,N)$.
The eigenstates of the Hubbard model in this configuration can be
represented in terms of an antisymmetrized product of independent
electron states in a Bloch basis, where the single particle energies
are given by
\begin{equation} \label{bloch}
\epsilon_j = -2t \cos \left(\frac{2 \pi j}{L} \right).
\end{equation}
In this expression,  the $j$'s are a
set of \textit{any} $L$ consecutive integers. The minimum energy
of this system  is thus obtained by
summing $N$ terms of the Bloch contributions (\ref{bloch}), resulting
in   the analytic form for
$E_{\rm{min}}(0,N)$:
\begin{equation} \label{minimum energy of maximal total spin states}
      \left\{
    \begin{array}{ll}
        \displaystyle -2t \sum_{j = -\frac{N}{2}+1}^{\frac{N}{2}} \cos \left(
        \frac{2 \pi j}{L} \right)
         = \displaystyle E_g \cos \left( \displaystyle \frac{ \pi}{L} \right),
              & \text{$N$ even} \\ \\
        \displaystyle -2t  \sum_{j = -\frac{N-1}{2}}^{\frac{N-1}{2}}
              \cos \left(
        \frac{2 \pi j}{L} \right) = \displaystyle E_g , & \text{$N$ odd,} \\ \\
    \end{array}
    \right.
\end{equation}
where $E_g$ is the energy given in Eq.\ (\ref{ground state
energy}). The form of Eq.\ (\ref{minimum energy of maximal total spin states})
shows an interesting simplification: if $N = L - 1$ is odd, then
$E_{\rm{min}}(0,N) = -2t$.  Interestingly, Trugman\cite{Trugman:1990} proved the general result
that the lowest bound energy of the ground state of the one dimensional Hubbard model in the atomic limit is
exactly $-2t(L - N)$, where $L - N$ is the number of holes. Therefore it follows that the ground state
energy of a one-hole system with an odd number of electrons must be $E_g = -2t$. Furthermore, there must exist precisely one state with that energy with maximal total spin.

\begin{figure}
    \includegraphics[width=80mm]{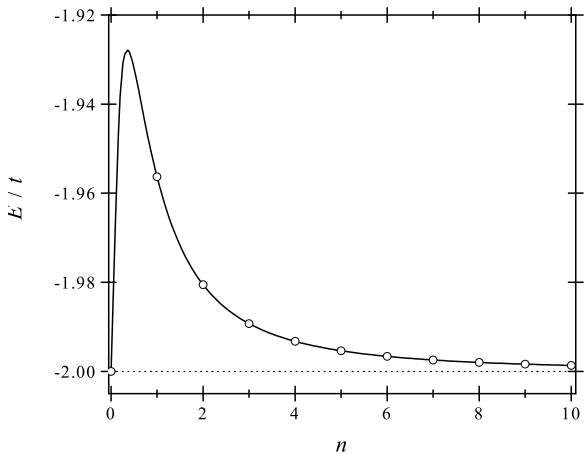}
    \caption{Plots of the exact ground state energy given in
Eq.\ (\ref{ground state energy}) (dashed line)
    and the lowest energy obtained using the Bethe ansatz charge momenta
    Eq.\ (\ref{one hole k_j})
    in Eq.\ (\ref{Bethe ansatz energy}) (solid line)
     for a system with  $N = L - 1 = 4n + 1$,
    as a function of $n$. The circles denote integer values of the variable $n$.}
    \label{Figure: One-Hole Comparison}
\end{figure}

We continue our examination of a one-hole system with the special case\cite{Sorella:1992} of $N = L - 1 = 4n +
1$ and $N_\downarrow = 2n$, where $n$ is an integer. In this case, Eq.\ (\ref{Bethe k_j}) becomes
\begin{equation} \label{one hole k_j}
    k_j(2n,2n+1) = \frac{2 \pi}{4n + 2} \left[j + \frac{n}{4n + 1} \right]
    ,
\end{equation}
where $j$ is an integer. We find the possible energies of the
system by summing over $j$ in the formula given in Eq.\ (\ref{Bethe ansatz
energy}); the lowest energy obtained is plotted in Fig.
\ref{Figure: One-Hole Comparison} as a function $n$. Inspection of this plot reveals that using the charge
momenta given in Eq.\ (\ref{one hole k_j}) to evaluate
Eq.\ (\ref{Bethe ansatz energy}) results in energies that
are \textit{above} the ground state energy of $-2t$,
except in the one-electron
case $(n = 0)$ and the thermodynamic limit $(n = \infty)$.
The failure of this example for systems of finite size, indicates that
Eq.\ (\ref{Bethe k_j}) should be reexamined.

\begin{figure}
\centering
\includegraphics[width=80mm]{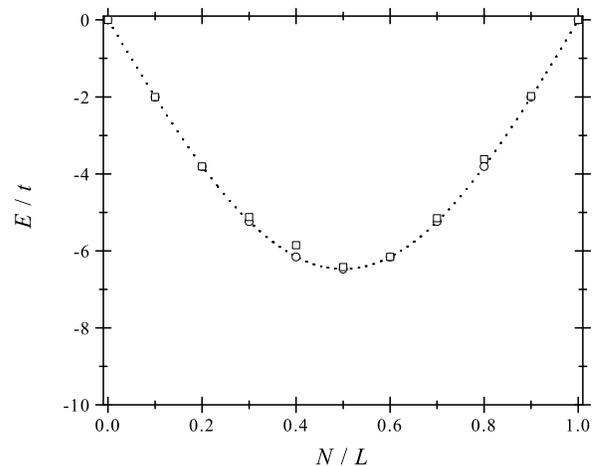}\\
\centering
\includegraphics[width=80mm]{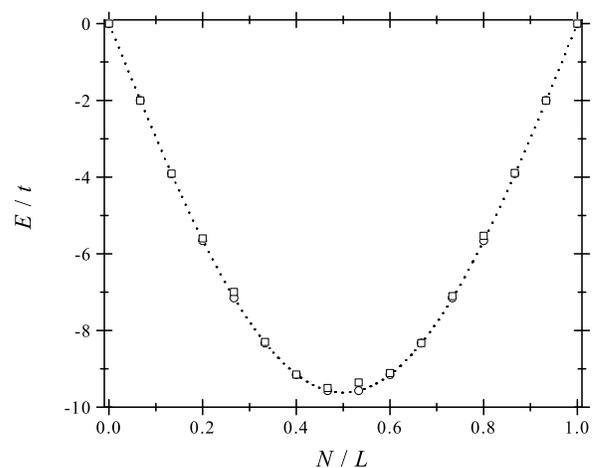}\\
\caption{Plots of the exact ground state energies  using
Eq.\ (\ref{ground state energy}) (circles) and energies obtained using the
unmodified Bethe ansatz solutions in Eqs.\ (\ref{Bethe ansatz
energy}) and (\ref{Bethe k_j}) (squares) at $u = \infty$ for $L =
10$ (\textit{top}) and $L = 15$ (\textit{bottom}) as a function of
the chain density $N / L$. The dotted line connecting the circles
is obtained by treating $N$ as a continuous variable.}
\label{Figure: Bethe Energy Comparisons}
\end{figure}

To address this problem, we recall that
the Hubbard Hamiltonian obeys the commutation relations
\begin{equation} \label{commutation relations}
    \left[ \mathcal{H} , \mathbf{S}^2 \right] =
    \left[ \mathcal{H}, S_z \right] =
    \left[ \mathcal{H}, S_{\pm} \right] = 0 ,
\end{equation}
where $\mathbf{S}^2$ and $S_z$ are the operators for total spin
and $z$-component of spin, and $S_{+}$ and $S_{-}$ are the spin
raising and lowering operators. Because of these relations, it is possible
to find
the energy eigenstates of the Hubbard Hamiltonian which are
also  eigenstates of
$\mathbf{S}^2$ and $S_z$ with quantum numbers $S$ and $M^{}_S$,
respectively.

The use of the charge momenta relations in Eq.\ (\ref{Bethe k_j}) does not take this spin symmetry into account.
In order to include this symmetry in the Bethe ansatz formulation,
we consider the relationships between systems having different spin configurations.
We assume that the energy eigenfunctions $\psi(N_\downarrow, N_\uparrow)$ are
known
\begin{equation} \label{eigenvalue equation M = N/2}
    \mathcal{H} \psi(N_\downarrow, N_\uparrow) =
    E \psi(N_\downarrow, N_\uparrow) ,
\end{equation}
and are simultaneously eigenfunctions of $\mathbf{S}^2$ with total spin
quantum number $S$. Applying the spin-raising operator $S_{+}$ to Eq.\
(\ref{eigenvalue equation M = N/2}), one of two things  occurs: either the
eigenfunction $\psi(N_\downarrow, N_\uparrow)$ is annihilated if
it has minimal total spin quantum number $S =
\frac{1}{2} \left( N_\uparrow - N_\downarrow \right)$, or we obtain the eigenvalue equation
\begin{equation} \label{eigenvalue equation M = N/2 - 1}
    \mathcal{H} \psi (N_\downarrow - 1, N_\uparrow + 1) = E \psi
    (N_\downarrow - 1, N_\uparrow + 1) ,
\end{equation}
where $E$ is the same energy that appears in Eq.\ (\ref{eigenvalue
equation M = N/2}). Therefore, the set of eigenvectors and eigenvalues that
solves Eq.\ (\ref{eigenvalue equation M = N/2 - 1}) is a subset of
the set of eigenvectors and eigenvalues that solves Eq.\ (\ref{eigenvalue
equation M = N/2}).   This process can be repeated to span all of the
possible related spin configurations corresponding to the same energy
eigenvalue $E(M,M') = E(N_\downarrow,N_\uparrow)$, where the possible
values of $(M,M')$ are determined from limits of the raising operations to be
$0 \le M \le N_\downarrow$, $N_\uparrow \le M' \le N$, and $M+M'=N$.

In determining the ground state energy of our system,
 we are not actually constructing the eigenstate wavefunctions, but
searching for  the minimum energy represented by Eq.\ (\ref{Bethe ansatz
energy}) through the use of the
 charge momenta in Eq.\ (\ref{Bethe k_j}).  The result of the spin raising operator analysis,
 suggests that Eq.\ (\ref{Bethe k_j}) should be modified to take the form:
\begin{equation} \label{ModBethek}
    k_j(N_\downarrow, N_\uparrow) = \left\{
    \begin{array}{ll}
        \displaystyle \frac{2 \pi I_j}{L} , & \text{$N$ even}
        \\ \\
        \displaystyle \frac{2 \pi}{L} \left( I_j + \frac{M}{2N}
        \right) , & \text{$N$ odd.}
    \end{array}
    \right.
\end{equation}
where $0 \le M \le N_\downarrow$.

With this modified result, we now minimize the system energy in Eq.\ (\ref{Bethe ansatz
energy}).
We write the charge momenta appearing in Eq.\ (\ref{Bethe ansatz
energy}) in a general form consistent with Eq.\ (\ref{ModBethek})
\begin{equation} \label{general k_j}
    k_j = \frac{2 \pi}{L} \left( j + j_0 \right) ,
\end{equation}
where the $j$'s are positive integers and $j_0$ is a real number.
With this choice of charge momenta $k_j$,
 Eq.\ (\ref{Bethe ansatz energy}) can be summed as a geometric series
resulting in the energy equation
\begin{align} \label{geometric series of cosines}
    E
    & =
    -2t \sum_{j = 1}^{N} \cos \left[ \frac{2 \pi}{L} \left(j + j_0
    \right) \right] \nonumber \\
    & =
    -2t \displaystyle \frac{\sin \left( \pi N / L \right)}{\sin
    \left( \pi / L \right)}
    \cos \left[ \frac{ \left( 2j_0 + N + 1 \right) \pi}{L} \right]
    .
\end{align}
This expression has a minimum when
\begin{equation} \label{j_0 minimum}
    j_0 = -\frac{N+1}{2}.
\end{equation}

Now we must check whether this mathematical minimum is consistent with the
modified Bethe ansatz charge momenta $ k_j(N_\downarrow, N_\uparrow)$ given
by Eq.\ (\ref{ModBethek}).   First consider the case of $N$  even.
For this case Eq.\ (\ref{j_0 minimum}) is a half-odd-integer and
 $I_j(M) = j + j_0$ must be a half-odd-integer, which is realized
when
$M$ is  odd. Since for all even $N>0$, there is at least one choice of odd
$M$ in the range $0 \le M \le N/2$, this case is consistent with the modified
Bethe ansatz solution.     Now consider the case of $N$ odd.
For this case, Eq.\ (\ref{j_0 minimum}) must be an integer and
$I_j(M)+ M/(2N) = j + j_0$ must be an integer, which is realized when
 $M = 0$.
 To summarize all of these possibilities
we conclude that
the charge momenta corresponding to the ground state can
be chosen using consecutive integers $j$ centered at the origin of the form
\begin{equation} \label{Bethe k_j ground state}
    k_j(N_\downarrow, N_\uparrow) = \left\{
    \begin{array}{ll}
        \displaystyle \frac{2 \pi}{L} \left( j + \frac{1}{2}
        \right) , & \text{$N$ even} \\ \\
        \displaystyle \frac{2 \pi j}{L} , & \text{$N$ odd} ,
    \end{array}
    \right. .
\end{equation}
Using these charge momenta, and minimizing Eq.\ (\ref{Bethe ansatz energy}), we obtain the ground state energy expression in Eq.\ (\ref{ground state energy}). This concludes our derivation.

\begin{table}
    \caption{Examples of minimum energies obtained using Eqs.\
    (\ref{Bethe ansatz energy}) and (\ref{k_j}) and particular
    choices of the $J_\alpha$ parameters. The ground state energies
    $E_{g}$ were found using Eq.\ (\ref{ground state energy}) and were verified by exact
    diagonalization.
    \label{Table: Bethe Ansatz Energies}}
    \centering
    \begin{tabular}{cccccc}
    \toprule
    \hspace{4mm} $L$ \hspace{4mm} & \hspace{2mm} $N$ \hspace{2mm} & $E^{}_{g}/t$ & $N_\downarrow$ $(N_\uparrow)$
    & \hspace{1mm} $J^{}_{\alpha}$ \hspace{1mm} & \hspace{1mm} $E_{\mathrm{min}}/t$ \hspace{1mm} \\
    \otoprule
    \multirow{5}{*}{$6$} 
        & \multirow{3}{*}{$4$}
            & \multirow{3}{*}{$-3.46410$}
                & \multirow{3}{*}{$2$ $(2)$} 
                    & $-\frac{1}{2} , \frac{1}{2}$ & $-3.00000$ \\
                    & & & & $-\frac{1}{2} , \frac{3}{2}$ & $-3.34607$ \\
                    & & & & $-\frac{1}{2} , \frac{5}{2}$ & $-3.46410$ \\
                \cline{2-6}
        & \multirow{2}{*}{$5$}
            & \multirow{2}{*}{$-2.00000$}
                & \multirow{2}{*}{$2$ $(3)$}
                    & $0, 1$ & $-1.95630$ \\
                    & & & & $-1, 1$ & $-2.00000$ \\
                \cline{1-6}
    \multirow{10}{*}{$7$}
        & \multirow{3}{*}{$4$}
            & \multirow{3}{*}{$-4.49396$}
                & \multirow{3}{*}{$2$ $(2)$}
                    & $-\frac{1}{2} , \frac{1}{2}$ & $-4.04892$ \\
                    & & & & $-\frac{1}{2}, \frac{3}{2}$ & $-4.38129$ \\
                    & & & & $-\frac{1}{2} , \frac{5}{2}$ & $-4.49396$ \\
                \cline{2-6}
        & \multirow{2}{*}{$5$}
            & \multirow{2}{*}{$-3.60388$}
                & \multirow{2}{*}{$2$ $(3)$}
                    & $0, 1$ & $-3.54596$ \\
                    & & & & $-1, 1$ & $-3.60388$ \\
                \cline{2-6}
        & \multirow{5}{*}{$6$}
            & \multirow{5}{*}{$-2.00000$}
                & $3$ $(3)$ & $-1, 0, 1$ & $-2.00000$ \\ \cline{4-6}
                & & & \multirow{4}{*}{$2$ $(4)$}
                    & $-\frac{1}{2} , \frac{1}{2}$ & $-1.80194$ \\
                    & & & & $-\frac{1}{2}, \frac{3}{2}$ & $-1.91115$ \\
                    & & & & $-\frac{1}{2}, \frac{5}{2}$ & $-1.97766$ \\
                    & & & & $-\frac{1}{2} , \frac{7}{2}$ & $-2.00000$ \\
                \bottomrule
    \end{tabular}
\end{table}

Additional confirmation of Eq.\ (\ref{ground
state energy}) was obtained by direct diagonalization of the Hubbard
Hamiltonian in the $u \rightarrow \infty$ limit.
 Furthermore, it is easy to show that Eq.\
(\ref{ground state energy}) is consistent with previously found
lower bounds to the ground state energy.\cite{Trugman:1990}
In Fig. \ref{Figure: Bethe Energy
Comparisons} a comparison between the lowest energies found using
Eqs. (\ref{Bethe ansatz energy}) and (\ref{Bethe k_j}), and the
energy found using Eq.\ (\ref{ground state energy}) is shown for
$L = 10$ and $L = 15$ at various lattice densities.  On this scale,
the error introduced by using Eq.\ (\ref{Bethe k_j}) is small but
non-trivial.

In addition to the analysis presented above, there
are other possible ways to find the ground state energies from the
Lieb-Wu equations. Instead of restricting the $J_\alpha$'s to be consecutive integers or
half-odd-integers centered around the origin as assumed in the
above derivation of Eq.\ (\ref{ground state energy}), it is also
 possible to find energies (including
the ground state energy) using Eqs.\ (\ref{Bethe ansatz
energy}) and (\ref{k_j}) but with non-consecutive $J_\alpha$'s.
In Table \ref{Table: Bethe Ansatz Energies} we list
several examples of this situation.
In all cases we have investigated,
the ground state energies determined in this way are
consistent with Eq.\ (\ref{ground state energy}).
One might hope that such an approach could generate excited states in addition
to the ground states of the system.  Unfortunately, there is no guarantee that
the  Bethe ansatz solutions are complete; in fact  for even $L$ it has been
shown\cite{Essler:1991,Essler:1992} that extensions are needed to
obtain a complete set of solutions.

\section{Numerical examples \label{illus}}

\begin{figure}
    \centering
    \includegraphics[width=80mm]{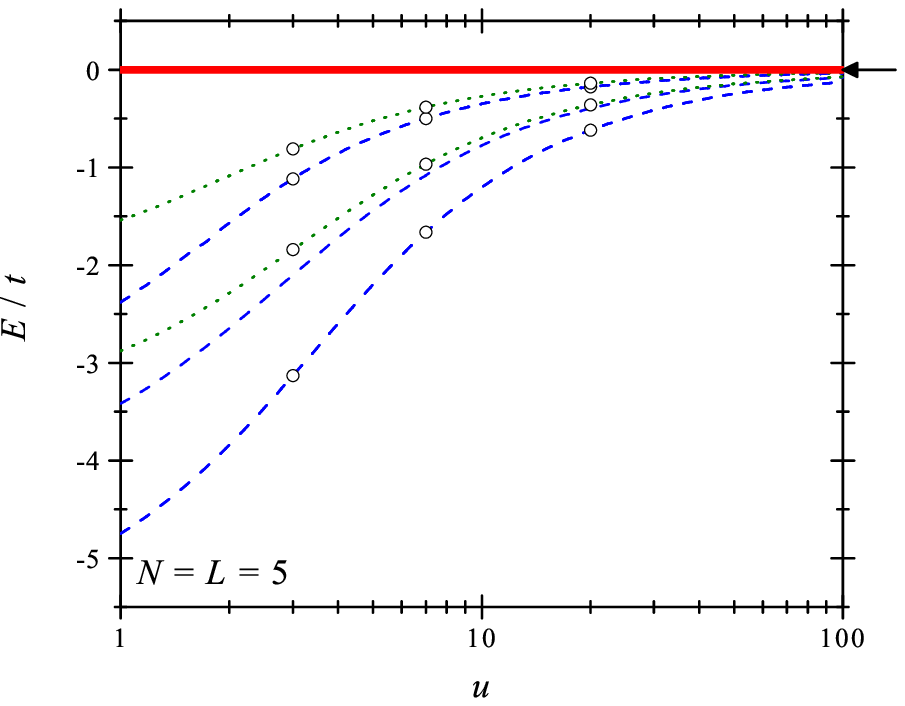}\\
    \centering
    \includegraphics[width=80mm]{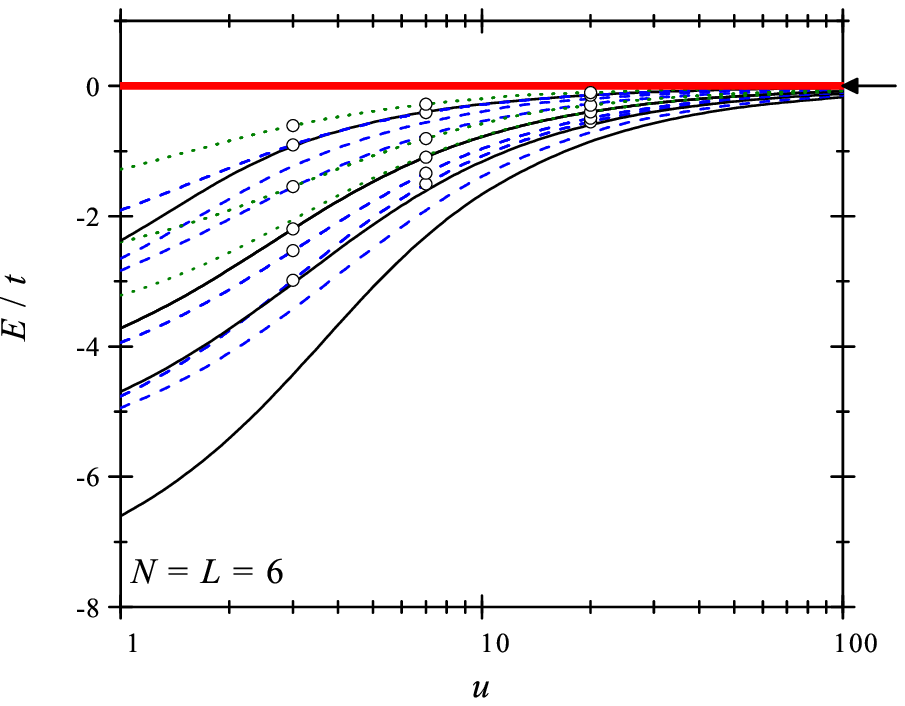}\\
    \caption{Plot of the lowest energies as a function of $u$ for (\textit{top}) $N = L = 5$ and (\textit{bottom}) $N = L = 6$. A thick (red) line represents a maximal spin $S = N/2$ state; a dotted (green) curve represents a $S = (N - 2)/2$ state; a dashed (blue) curve represents a $S = (N-4)/2$ state. In the bottom plot, a solid (black) curve represents a $S = (N-6)/2 = 0$ state. Symbols denote a doubly degenerate energy. Arrows mark the energies obtained by Eq.\ (\ref{ground state energy}).}
    \label{Figure: Half Filling}
\end{figure}
\begin{figure}
    \centering
    \includegraphics[width=80mm]{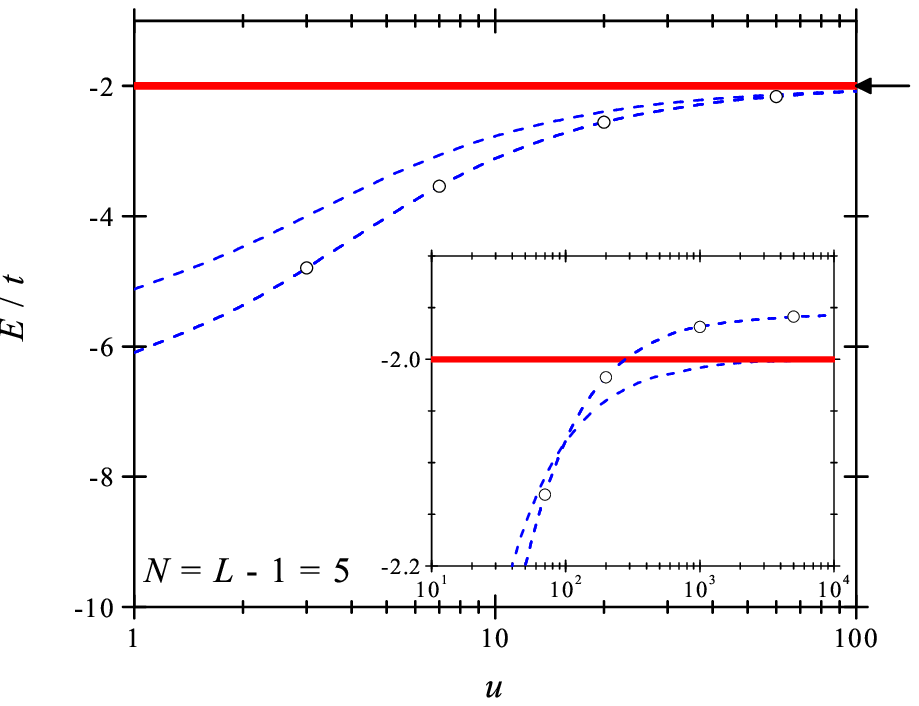}\\
    \centering
    \includegraphics[width=80mm]{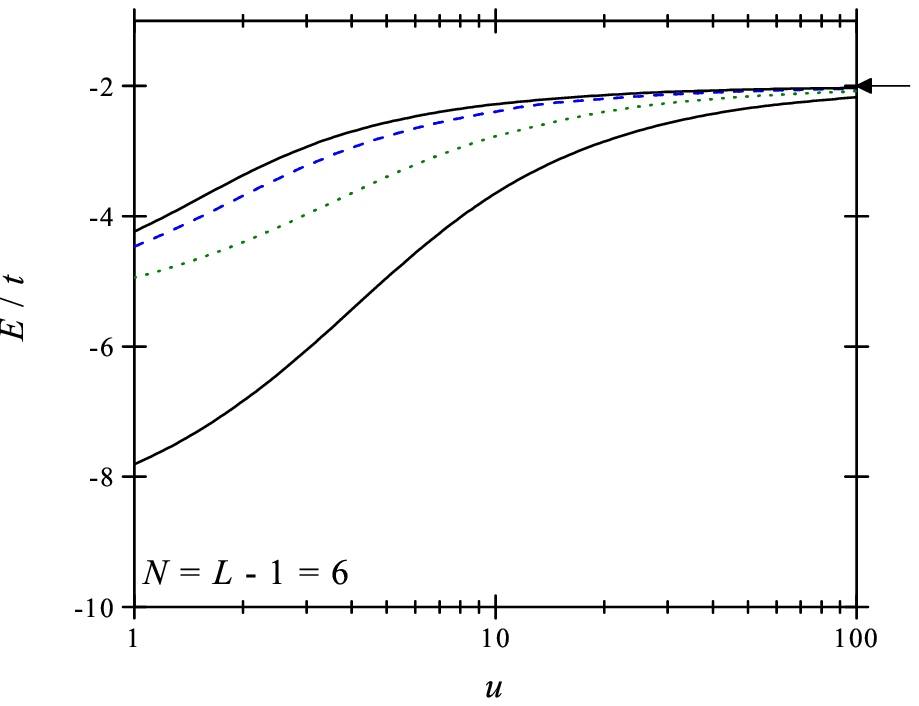}\\
    \caption{Plot of the lowest energies as a function of $u$ for (\textit{top}) $N = L - 1 = 5$ and (\textit{bottom}) $N = L - 1 = 6$, using the same line (color) style as Figure \ref{Figure: Half Filling}. The degeneracy of the ground state in the atomic limit is given by Eqs.\ (\ref{one hole degeneracy}) (\textit{top}) and (\ref{one hole degeneracy even}) (\textit{bottom}).}
    \label{Figure: One Hole}
\end{figure}
\begin{figure}
    \centering
    \includegraphics[width=80mm]{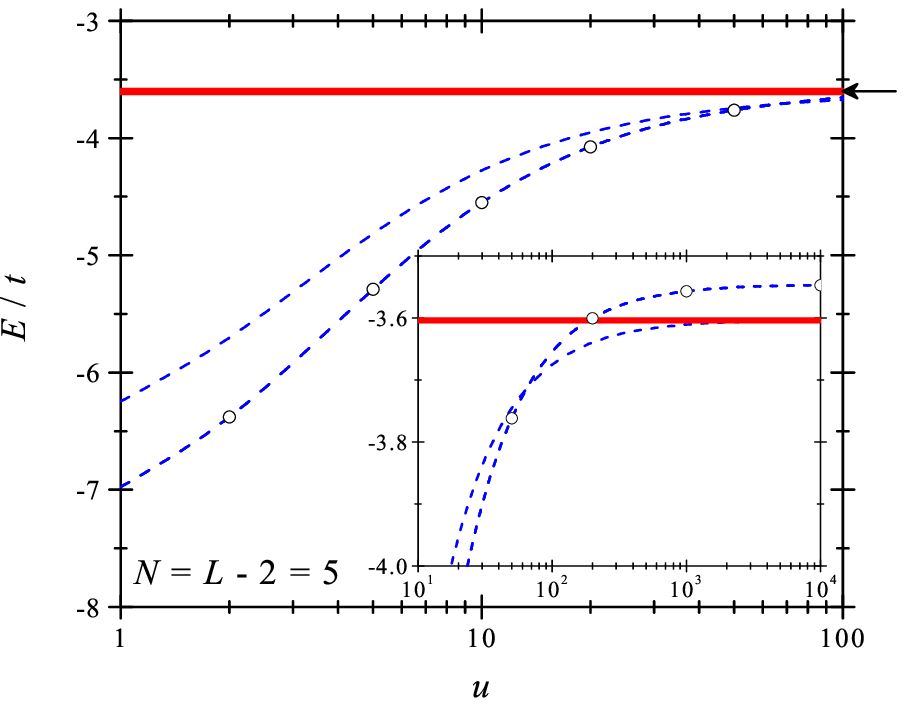}\\
    \centering
    \includegraphics[width=80mm]{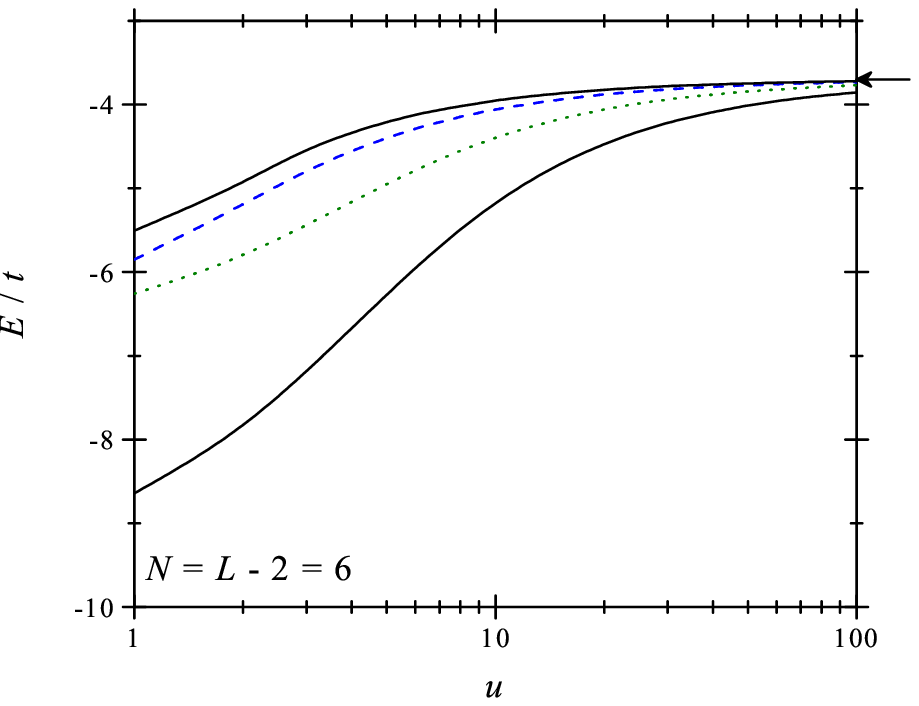}\\
    \caption{Plot of the lowest energies as a function of $u$ for (\textit{top}) $N = L - 2 = 5$ and (\textit{bottom}) $N = L - 2 = 6$, using the same line (color) style as Figure \ref{Figure: Half Filling}.}
    \label{Figure: Two Holes}
\end{figure}

In order to understand the results presented here, we consider some specific examples. Throughout this section, we assume, without loss of generality, that $N_\uparrow - N_\downarrow = 0$ ($+1$) if $N$ is even (odd). For ground states in the atomic limit, the probability of two electrons occupying the same lattice site vanishes. Therefore, there are only two configurations for a lattice site: vacant (henceforth referred to as a \textit{hole}) or singly occupied. In general, for a system of $n$ holes, the ground state is highly degenerate, with a degeneracy denoted by $d^{(n)}$.

We begin our examination with the simplest case (\textit{half-filling}) where the number of holes is zero. For any half-filled system in the atomic limit, the ground states have energy $E_g = 0$. The degeneracy of the ground states is the number of possible ways of filling $N = L$ holes with $N$ indistinguishable spin-$1/2$ particles
\begin{equation} \label{n = 0 degeneracy}
    d^{(0)} = \frac{N!}{N_\uparrow! N_\downarrow!}.
\end{equation}

In Figure \ref{Figure: Half Filling}, the energies of these $d^{(0)}$ states, that become the ground states in the atomic limit, are plotted as a function of $u$ for two small half-filled systems: $N = 5$ and $N = 6$. Results were obtained by exact diagonalization. In either case, the degeneracy of the ground state in the atomic limit is given by Eq.\ (\ref{n = 0 degeneracy}): for $N = 5$, $d^{(0)} = 10$; for $N = 6$, $d^{(0)} = 20$. As predicted by Eq.\ (\ref{minimum energy of maximal total spin states}), in both cases, since $E_g = 0$, one of the ground states has maximal total spin. This state is easily distinguished by having an energy that is independent of the coupling parameter $u$.

Next, we introduce one hole and examine its effect. In Figure \ref{Figure: One Hole}, the lowest energies of two small one-hole systems, $N = L - 1= 5$ and $N = 6$, are plotted as a function of $u$. According to Eq.\ (\ref{ground state energy}), the energy of the ground states in the atomic limit for \textit{any} one hole system is $E_g = -2t$. In order to determine the degeneracy, we use perturbation theory, taking $\mathcal{H}_{\mathrm{int}}$ as the unperturbed Hamiltonian. The \textit{unperturbed} ground states are degenerate and have zero energy
\begin{equation} \label{unperturbed one hole ground state}
    \mathcal{H}_{\mathrm{int}} (U \rightarrow \infty) | i \rangle = E^{(0)}_i | i \rangle = 0 .
\end{equation}
This unperturbed degeneracy $D$ is given by the number of possible ways of filling $L$ holes with $N = L - 1$ indistinguishable spin-$1/2$ particles
\begin{equation} \label{degeneracy of unperturbed one hole}
    D = \frac{(N + 1)!}{N_\uparrow! N_\downarrow!} .
\end{equation}

Treating the hopping term $\mathcal{H}_{\mathrm{hop}}$ as the perturbation, the first order correction to the ground state energy is the lowest eigenvalue of the matrix
\begin{equation} \label{one-hole perturbation}
    \mathcal{W}^{(1)}_{k \ell} = \langle k | \mathcal{H}_{\mathrm{hop}} (t) | \ell \rangle .
\end{equation}
In the atomic limit, higher order corrections to the ground state energy vanish; the lowest energy eigenstates of the perturbation matrix in Eq.\ (\ref{one-hole perturbation})  are the exact ground states in the atomic limit. Furthermore, we observe that this perturbation matrix \textit{can} be put in block diagonal form, by appropriate ordering of the states from Eq.\ (\ref{unperturbed one hole ground state}). To prove this assertion we examine more closely these states.

\begin{figure}
    \includegraphics[width=80mm]{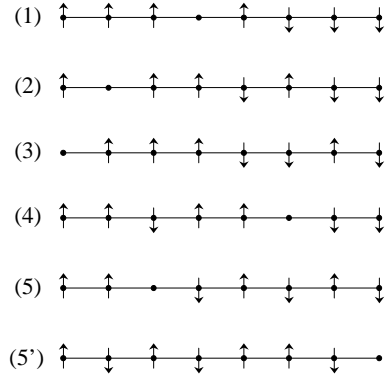}
    \caption{Possible one-dimensional spin arrangements for $N = L - 1 = 7$ and $N_\downarrow = 3$. Arrangements $(1)$ through $(4)$ each represent a different spin configuration for this system; arrangements $(5)$ and $(5')$ represent the fifth and final spin configuration. Arrangements $(1)$ through $(4)$ are disconnected from one another and from arrangements $(5)$ and $(5')$; arrangements $(5)$ and $(5')$ are connected to one another.}
    \label{Figure: Disconnected_States}
\end{figure}

In Figure \ref{Figure: Disconnected_States} we show several unperturbed ground states defined in Eq.\ (\ref{unperturbed one hole ground state}) for the case $N = L - 1 = 7$ and $N_\downarrow = 3$. With periodic boundary conditions, the state depicted in $(5')$ can be obtained from arrangement $(5)$ by repeatedly moving electrons one at a time to nearest neighbor sites with the provision that all intermediate states contain no doubly occupied lattice sites; the same cannot be said for states $(1)$ through $(4)$. Therefore, states $(5)$ and $(5')$ are said to be \textit{connected} to each other and \textit{disconnected} from states $(1)$ through $(4)$. In fact, states $(1)$ through $(5)$ represent the five distinct \textit{spin configurations} for that system. To indicate that the unperturbed ground states can be sorted into different spin configurations, we write them as $| \alpha, i_\alpha \rangle$, where the integer $\alpha$ denotes to which spin configuration the state belongs, and $i_\alpha$ is an arbitrary state label. Since states in different spin configurations are disconnected,
\begin{equation}
    \langle \alpha ', i'_{\alpha '} | \mathcal{H}_{\mathrm{hop}} | \alpha, i_\alpha \rangle = 0, \;\;\; \mathrm{if} \; \alpha \neq \alpha' .
\end{equation}
Therefore, the first order perturbation matrix is block diagonal, with the number of blocks being the number of distinct spin configurations for that system.

In general, the number of blocks is a complicated function that depends upon our choice of $N$ and $L$. We treat the simplest case, when $N$ is odd, first. Then, it is easy to show that the number of connected arrangements in each spin configuration is
\begin{equation} \label{C}
    \mathcal{C} = NL = N(N+1);
\end{equation}
this expression is valid provided that $N_\sigma \neq 0$ and $N_\uparrow/N_\downarrow$ and its inverse are not integers for $N_\sigma > 1$. With the assumption that $N_\uparrow = N_\downarrow + 1$, this expression is necessarily valid for an odd number of electrons. Therefore, the number of distinct spin configurations is
\begin{equation} \label{one hole degeneracy}
    \frac{D}{\mathcal{C}} = \frac{(N-1)!}{N_\uparrow! N_\downarrow!} = d^{(1)} ;
\end{equation}
the last equality will be shown below.

Returning to Eq.\ (\ref{one-hole perturbation}), we factor the first-order correction as
\begin{equation} \label{one-hole perturbation matrix}
    \mathcal{W}^{(1)}_{k \ell} = -2t \langle k | \mathcal{H}_{\mathrm{hop}} \left( - \tfrac{1}{2} \right) | \ell \rangle;
\end{equation}
the lowest eigenvalue of $\mathcal{W}_{(1)}$ corresponds to the maximum eigenvalue of the dimensionless operator $\mathcal{H}_{\mathrm{hop}} (-\tfrac{1}{2})$. By appropriately choosing the unperturbed basis defined in Eq.\ (\ref{unperturbed one hole ground state}), it can be shown that if $N$ is odd, then each row and column of $\mathcal{H}_{\mathrm{hop}} (-\tfrac{1}{2})$ consists of only two nonzero elements, which have the value $\frac{1}{2}$. Therefore,
\begin{equation}
    \langle k | \mathcal{H}_{\mathrm{hop}} \left( - \tfrac{1}{2} \right) | \ell \rangle \geq 0,
\end{equation}
and the sum of the elements in every row or column is 1
\begin{equation} \label{stochastic 1}
    \sum_{| k \rangle} \langle k | \mathcal{H}_{\mathrm{hop}} \left( -\tfrac{1}{2} \right) | \ell \rangle = 1
\end{equation}
and
\begin{equation} \label{stochastic 2}
    \sum_{| \ell \rangle} \langle k | \mathcal{H}_{\mathrm{hop}} \left( -\tfrac{1}{2} \right) | \ell \rangle = 1 .
\end{equation}
Nonnegative matrices that satisfy either Eq.\ (\ref{stochastic 1}) or Eq.\ (\ref{stochastic 2}) are known as \textit{stochastic matrices}\cite{Berman:1979} and are well studied. These matrices describe the transitions of a Markov chain; their elements are the transition probabilities that a system will jump from one state to another. Equations (\ref{stochastic 1}) and (\ref{stochastic 2}) state that the total probability of transition is unity. By the Perron-Frobenius theorem\cite{Berman:1979}, the maximal eigenvalue of these matrices is always 1, with the corresponding unnormalized eigenstate being the unity vector, whose elements are all 1.

Since each block is a stochastic matrix, we conclude that the ground state is $d^{(1)}$-fold degenerate. Furthermore, from Eq.\ (\ref{one-hole perturbation matrix}), the energy of the ground state is $-2t$, in agreement with Eq.\ (\ref{ground state energy}). For the case $N = L - 1 = 5$ and $N_\uparrow = N_\downarrow + 1 = 3$, Eq.\ (\ref{one hole degeneracy}) predicts that in the atomic limit, $d^{(1)} = 2$, which is confirmed in Figure \ref{Figure: One Hole}. For small values of $u > 0$, the ground state energy is two-fold degenerate. As we increase $u$, an energy crossing occurs in the vicinity $u \approx 100$ and a different state becomes the ground state. As $u \rightarrow \infty$, this state converges to the energy of the lowest-energy maximal spin state, which is $-2t$. In the atomic limit, the higher energy states plotted in Figure \ref{Figure: One Hole} have energy $-1.95630 t$; this result corresponds to the case of $n = 1$ in Figure \ref{Figure: One-Hole Comparison}, which was obtained by using the charge momenta in Eq.\ (\ref{one hole k_j}). This was identified as the ground state in the atomic limit in Ref. \onlinecite{Sorella:1992}.

For an even number of electrons, the situation is slightly more complicated, since Eq.\ (\ref{C}) is not valid for every spin configuration. Particular spin configurations may have additional periodicities that decrease the number of connected states in that configuration. However, with our assumption that $N_\uparrow = N_\downarrow$, we find that if $N_\sigma$ is prime, then there is only one spin configuration which does not obey Eq.\ (\ref{C}). This configuration contains all the states defined in Eq.\ (\ref{unperturbed one hole ground state}) where every pair of electron neighbors have opposite spin (a one-hole antiferromagnetic chain). Obviously, the number of connected states in this spin configuration is $2L$. Therefore, the degeneracy is
\begin{equation} \label{one hole degeneracy even}
    d^{(1)} = \frac{D-2L}{C} + 1 = \frac{(N-1)!}{N_\uparrow! N_\downarrow!} - \frac{2}{N} + 1 .
\end{equation}
For the case of $N = L - 1 = 6$ and $N_\sigma = 3$ (shown in Figure \ref{Figure: One Hole}), $d^{(1)} = 4$.

The case of two holes in the atomic limit is much more complex, and the methods used for one hole are not applicable. In Figure \ref{Figure: Two Holes}, the lowest energies of two small two-hole systems, $N = L - 2 = 5$ and $N = 6$, are plotted as a function of $u$. In general, the low-lying energies in Figure \ref{Figure: Two Holes} show a striking similarity to the one-hole energies plotted in Figure \ref{Figure: One Hole}. Again, we find that in the case $N = L - 2 = 5$, there is an exchange of ground states in the vicinity $u \approx 100$.

\section{Discussion and conclusions \label{discuss}}

The form of the ground state energy given by Eq.\ (\ref{ground state energy})
and its derivation provides additional insight into the nature of the
eigenstates of the one-dimensional Hubbard model in the atomic limit.
The form of the energy Eq.\ (\ref{ground state energy}) shows that there is
an electron-hole symmetry in the ground state energy such that the energy
of  a system with $N$ electrons is identical to a system with $L-N$ electrons,
corresponding to $L-N$ and $N$ holes, respectively.
In the course of  deriving  Eq.\ (\ref{ground state energy}), we used Eq.\
(\ref{minimum energy of maximal total spin states}) to show that
for an odd number of electrons $N$, or at half-filling ($N = L$),
the maximal total electron spin configuration has the ground state
energy given in Eq.\ (\ref{ground state energy}) for any system size $L \ge N$.
The conclusion that for an odd number of electrons in the atomic limit, exactly one of the ground states
(not counting the trivial $2S_{\mathrm{max}} + 1 = N + 1$
degeneracy) has maximal total spin has also been discussed
by previous authors\cite{Aizenman:1990}. If $N$ is even (except for the case of half-filling), the minimum energy of a state with maximal total spin is above the ground state energy, as shown in
Eq.\ (\ref{minimum energy of maximal total spin states}).

Numerical results illustrate the asymptotic behavior of the energies of the ground state and low-lying excited states as $u \rightarrow \infty$. In all the cases presented, the absolute error per particle of the $u = \infty$ energy given in Eq.\ (\ref{ground state energy}) to the exact energy of the ground state at $u = 100$ is less than $0.03t$ per particle.

Due to its rich structure and relative simplicity, there is an
impressive literature devoted to solutions of the one-dimensional
Hubbard model. To the best of our knowledge, Eq.\ (\ref{ground state energy}) and
the derivation presented here has not appeared in the previous literature. On the other hand, there are
closely related works. For example, Kotrla\cite{Kotrla:1990} extended
the approach of Caspers and Iske\cite{Caspers:1989} to consider the
$u \rightarrow \infty$ limit of the one dimensional Hubbard model
from the view point of enumeration of all possible single occupancy
states of the system. The analysis of the minimum energy configuration
results in an expression that is equivalent to our
Eq.\ (\ref{ground state energy}) although the explicit analytic form is
not given. Another related result is by Schadschneider
\cite{Schadschneider:1995}, who augments the electron hopping term of
the original Hubbard model in Eq.\ (\ref{Hubbard Hamiltonian}) with
a bond-charge interaction with strength parameter $X$. When $X = t$, the
number of doubly occupied sites becomes a conserved quantity. The energies
of the modified Hamiltonian are determined. In the limit as
$u \rightarrow \infty$, the ground state energy of that model is the
same as Eq.\ (\ref{ground state energy}). However, this is for $X = t, u
\rightarrow \infty$, whereas our result is for $X = 0, u \rightarrow
\infty$. Recently, Kumar\cite{Kumar:2009} considered
the fixed boundary solutions of a one-dimensional Hubbard system in the infinite $u$ limit.
For this case, the Lieb-Wu analysis is not applicable and the energy
spectrum is quite different.  Kumar was able to
find the analogue of Eq.\ (\ref{ground state energy}) for the fixed boundary case.

More detailed analysis has been devoted to case where $L$ is even which
for periodic boundary conditions allows for bipartite symmetry.\cite{Lieb:2003} Essler
\textit{et al.}\cite{Essler:1991,Essler:1992} derive a method for finding all of
the energy eigenstates by augmenting the Bethe ansatz using generators
associated with the SO(4) symmetry of the system.
Their results are presumably consistent with those in this paper,
though they do not explicitly evaluate their equations in the $u
\rightarrow \infty$ limit. Lieb and Wu\cite{Lieb:2003} and Goldbaum\cite{Goldbaum:2005}
prove the existence of ground state solutions to the Bethe ansatz equations for the restricted case
of even $N = L$ and odd $N_\sigma = N/2$, and show that the ground state is non-degenerate. It should be noted that in this case, the ground state is unique only for finite values of $u$; in general, at $u = \infty$ the ground state is degenerate. Figure \ref{Figure: Half Filling} illustrates this for the case $N = L = 6$: for $u < \infty$ the ground
state is nondegenerate and has $S = 0$; at $u=\infty$, $d^{0} = 20$.

In summary, we have derived an expression for the ground state
energy of a Hubbard ring in the atomic limit for even and odd
integer $L$. This expression agrees with exact diagonalization
energies obtained for several small systems, and is consistent
with limiting results reported in the literature.\cite{Shiba:1972,
Trugman:1990}

\bibliographystyle{apsrev}
\bibliography{Bibliography_v2.0}

\begin{thebibliography}{23}
\expandafter\ifx\csname natexlab\endcsname\relax\def\natexlab#1{#1}\fi
\expandafter\ifx\csname bibnamefont\endcsname\relax
  \def\bibnamefont#1{#1}\fi
\expandafter\ifx\csname bibfnamefont\endcsname\relax
  \def\bibfnamefont#1{#1}\fi
\expandafter\ifx\csname citenamefont\endcsname\relax
  \def\citenamefont#1{#1}\fi
\expandafter\ifx\csname url\endcsname\relax
  \def\url#1{\texttt{#1}}\fi
\expandafter\ifx\csname urlprefix\endcsname\relax\def\urlprefix{URL }\fi
\providecommand{\bibinfo}[2]{#2}
\providecommand{\eprint}[2][]{\url{#2}}

\bibitem[{\citenamefont{Hubbard}(1963)}]{Hubbard:1963}
\bibinfo{author}{\bibfnamefont{J.}~\bibnamefont{Hubbard}},
  \bibinfo{journal}{Proc. R. Soc. (London) A} \textbf{\bibinfo{volume}{276}},
  \bibinfo{pages}{238} (\bibinfo{year}{1963}).

\bibitem[{\citenamefont{Marder}(2001)}]{Marder:2001}
\bibinfo{author}{\bibfnamefont{M.}~\bibnamefont{Marder}},
  \emph{\bibinfo{title}{{C}ondensed {M}atter {P}hysics}}
  (\bibinfo{publisher}{{W}iley-{I}nterscience}, \bibinfo{year}{2001}).

\bibitem[{\citenamefont{Lieb and Wu}(1968)}]{Lieb:1968}
\bibinfo{author}{\bibfnamefont{E.~H.} \bibnamefont{Lieb}} \bibnamefont{and}
  \bibinfo{author}{\bibfnamefont{F.~Y.} \bibnamefont{Wu}},
  \bibinfo{journal}{Phys. Rev. Lett.} \textbf{\bibinfo{volume}{20}},
  \bibinfo{pages}{1445} (\bibinfo{year}{1968}).

\bibitem[{\citenamefont{Lieb and Wu}(2003)}]{Lieb:2003}
\bibinfo{author}{\bibfnamefont{E.~H.} \bibnamefont{Lieb}} \bibnamefont{and}
  \bibinfo{author}{\bibfnamefont{F.~Y.} \bibnamefont{Wu}},
  \bibinfo{journal}{Physica A} \textbf{\bibinfo{volume}{321}},
  \bibinfo{pages}{1} (\bibinfo{year}{2003}),
  \bibinfo{note}{arXiv.org:cond-mat/0207529}.

\bibitem[{\citenamefont{Bethe}(1931)}]{Bethe:1931}
\bibinfo{author}{\bibfnamefont{H.~A.} \bibnamefont{Bethe}},
  \bibinfo{journal}{Z. Phys.} \textbf{\bibinfo{volume}{71}},
  \bibinfo{pages}{205} (\bibinfo{year}{1931}).

\bibitem[{\citenamefont{Essler et~al.}(1991)\citenamefont{Essler, Korepin, and
  Schoutens}}]{Essler:1991}
\bibinfo{author}{\bibfnamefont{F.~H.~L.} \bibnamefont{Essler}},
  \bibinfo{author}{\bibfnamefont{V.~E.} \bibnamefont{Korepin}},
  \bibnamefont{and}
  \bibinfo{author}{\bibfnamefont{K.}~\bibnamefont{Schoutens}},
  \bibinfo{journal}{Phys. Rev. Lett.} \textbf{\bibinfo{volume}{67}},
  \bibinfo{pages}{3848} (\bibinfo{year}{1991}).

\bibitem[{\citenamefont{Essler et~al.}(1992)\citenamefont{Essler, Korepin, and
  Schoutens}}]{Essler:1992}
\bibinfo{author}{\bibfnamefont{F.~H.~L.} \bibnamefont{Essler}},
  \bibinfo{author}{\bibfnamefont{V.~E.} \bibnamefont{Korepin}},
  \bibnamefont{and}
  \bibinfo{author}{\bibfnamefont{K.}~\bibnamefont{Schoutens}},
  \bibinfo{journal}{Nuclear Physics B} \textbf{\bibinfo{volume}{384}},
  \bibinfo{pages}{431} (\bibinfo{year}{1992}).

\bibitem[{\citenamefont{Goldbaum}(2005)}]{Goldbaum:2005}
\bibinfo{author}{\bibfnamefont{P.~S.} \bibnamefont{Goldbaum}},
  \bibinfo{journal}{Commun. Math. Phys.} \textbf{\bibinfo{volume}{358}},
  \bibinfo{pages}{317} (\bibinfo{year}{2005}).

\bibitem[{\citenamefont{Nauciel-Bloch and Eggarter}(1979)}]{Eggarter:1979}
\bibinfo{author}{\bibfnamefont{M.}~\bibnamefont{Nauciel-Bloch}}
  \bibnamefont{and} \bibinfo{author}{\bibfnamefont{T.~P.}
  \bibnamefont{Eggarter}}, \bibinfo{journal}{Phys. Rev. A}
  \textbf{\bibinfo{volume}{19}}, \bibinfo{pages}{1862} (\bibinfo{year}{1979}).

\bibitem[{\citenamefont{Woynarovich}(1982)}]{WoynarovichIbid:1982}
\bibinfo{author}{\bibfnamefont{F.}~\bibnamefont{Woynarovich}},
  \bibinfo{journal}{J. Phys. C: Solid State Phys.}
  \textbf{\bibinfo{volume}{15}}, \bibinfo{pages}{85} (\bibinfo{year}{1982}),
  \bibinfo{note}{\textit{ibid}, \textbf{15}, 97 (1982)}.

\bibitem[{\citenamefont{Ogata and Shiba}(1990)}]{Ogata:1990}
\bibinfo{author}{\bibfnamefont{M.}~\bibnamefont{Ogata}} \bibnamefont{and}
  \bibinfo{author}{\bibfnamefont{H.}~\bibnamefont{Shiba}},
  \bibinfo{journal}{Phys. Rev. B} \textbf{\bibinfo{volume}{41}},
  \bibinfo{pages}{2326} (\bibinfo{year}{1990}).

\bibitem[{\citenamefont{Caspers and Iske}(1989)}]{Caspers:1989}
\bibinfo{author}{\bibfnamefont{W.~J.} \bibnamefont{Caspers}} \bibnamefont{and}
  \bibinfo{author}{\bibfnamefont{P.~O.} \bibnamefont{Iske}},
  \bibinfo{journal}{Physica A} \textbf{\bibinfo{volume}{157}},
  \bibinfo{pages}{1033} (\bibinfo{year}{1989}).

\bibitem[{\citenamefont{Kotrla}(1990)}]{Kotrla:1990}
\bibinfo{author}{\bibfnamefont{M.}~\bibnamefont{Kotrla}},
  \bibinfo{journal}{Physics Letters A} \textbf{\bibinfo{volume}{145}},
  \bibinfo{pages}{33} (\bibinfo{year}{1990}).

\bibitem[{\citenamefont{Schadschneider}(1995)}]{Schadschneider:1995}
\bibinfo{author}{\bibfnamefont{A.}~\bibnamefont{Schadschneider}},
  \bibinfo{journal}{Phys. Rev. B} \textbf{\bibinfo{volume}{51}},
  \bibinfo{pages}{10386} (\bibinfo{year}{1995}).

\bibitem[{\citenamefont{Kumar}(2009)}]{Kumar:2009}
\bibinfo{author}{\bibfnamefont{B.}~\bibnamefont{Kumar}},
  \bibinfo{journal}{Phys. Rev. B} \textbf{\bibinfo{volume}{79}},
  \bibinfo{pages}{155121} (\bibinfo{year}{2009}).

\bibitem[{\citenamefont{Shiba}(1972)}]{Shiba:1972}
\bibinfo{author}{\bibfnamefont{H.}~\bibnamefont{Shiba}},
  \bibinfo{journal}{Phys. Rev. B} \textbf{\bibinfo{volume}{6}},
  \bibinfo{pages}{930} (\bibinfo{year}{1972}).

\bibitem[{\citenamefont{Doucot and Wen}(1989)}]{Doucot:1989}
\bibinfo{author}{\bibfnamefont{B.}~\bibnamefont{Doucot}} \bibnamefont{and}
  \bibinfo{author}{\bibfnamefont{X.~G.} \bibnamefont{Wen}},
  \bibinfo{journal}{Phys. Rev. B} \textbf{\bibinfo{volume}{40}},
  \bibinfo{pages}{2719} (\bibinfo{year}{1989}).

\bibitem[{\citenamefont{Mielke}(1991)}]{Mielke:1991}
\bibinfo{author}{\bibfnamefont{A.}~\bibnamefont{Mielke}}, \bibinfo{journal}{J.
  Stat. Phys.} \textbf{\bibinfo{volume}{62}}, \bibinfo{pages}{509}
  (\bibinfo{year}{1991}).

\bibitem[{\citenamefont{Yang}(1967)}]{Yang:1967}
\bibinfo{author}{\bibfnamefont{C.~N.} \bibnamefont{Yang}},
  \bibinfo{journal}{Phys. Rev. Lett.} \textbf{\bibinfo{volume}{19}},
  \bibinfo{pages}{1312} (\bibinfo{year}{1967}).

\bibitem[{\citenamefont{Trugman}(1990)}]{Trugman:1990}
\bibinfo{author}{\bibfnamefont{S.~A.} \bibnamefont{Trugman}},
  \bibinfo{journal}{Phys. Rev. B} \textbf{\bibinfo{volume}{42}},
  \bibinfo{pages}{6612} (\bibinfo{year}{1990}).

\bibitem[{\citenamefont{Sorella and Parola}(1992)}]{Sorella:1992}
\bibinfo{author}{\bibfnamefont{S.}~\bibnamefont{Sorella}} \bibnamefont{and}
  \bibinfo{author}{\bibfnamefont{A.}~\bibnamefont{Parola}},
  \bibinfo{journal}{J. Phys.: Condens. Matter} \textbf{\bibinfo{volume}{4}},
  \bibinfo{pages}{3589} (\bibinfo{year}{1992}).

\bibitem[{\citenamefont{Berman and Plemmons}(1979)}]{Berman:1979}
\bibinfo{author}{\bibfnamefont{A.}~\bibnamefont{Berman}} \bibnamefont{and}
  \bibinfo{author}{\bibfnamefont{R.~J.} \bibnamefont{Plemmons}},
  \emph{\bibinfo{title}{{N}onnegative {M}atrices in the {M}athematical
  {S}ciences}} (\bibinfo{publisher}{{A}cademic {P}ress}, \bibinfo{year}{1979}).

\bibitem[{\citenamefont{Aizenman and Lieb}(1990)}]{Aizenman:1990}
\bibinfo{author}{\bibfnamefont{M.}~\bibnamefont{Aizenman}} \bibnamefont{and}
  \bibinfo{author}{\bibfnamefont{E.~H.} \bibnamefont{Lieb}},
  \bibinfo{journal}{Phys. Rev. Lett.} \textbf{\bibinfo{volume}{65}},
  \bibinfo{pages}{1470} (\bibinfo{year}{1990}).

\end{thebibliography}

\end{document}